\documentclass{article}
\usepackage[utf8]{inputenc}
\usepackage{amsmath}
\usepackage{graphicx}
\usepackage{setspace}
\usepackage[section]{placeins}
\usepackage{biblatex}

\title{A framework of population dynamics from first principles}
\date{}
\author{Gui Araujo}

\begin{document}

\maketitle
\tableofcontents

\newpage
\section{Introduction}

The aim of this manuscript is to contain the arguments and define the theoretical objects for building a general framework to model population dynamics from the ground up, relying mainly on the probabilistic landscapes defining the dynamics instead of the context-dependent physical specification of systems. I intend to keep updating and correcting this manuscript. The goal is for all the different parts to be able to communicate with each other and for models to be directly comparable and to maintain an explicit connection to the first principles sustaining them. This modeling paradigm will stem from a Bayesian perspective on model definition and interpretation and will be primarily concerned with ecological and evolutionary processes. Populations are considered to be abstract collections of elements that relate in the same ways, and the laws of motion ultimately depend on relational properties of elements, at first irrespective of their constitution. The states of populations are taken to be their spatial densities, the fundamental quantities shaping the dynamics of their interactions.

Organization of the manuscript: 2) a simple Bayesian motivation to introduce the way I think of probabilities and stochasticity. 3) Defining reaction networks as the language of model-building in terms of local relations. 4) Defining the fundamental description of the system as stochastic and governed by a master equation of Markov jump processes. 5) Deriving the law of transition between states that will drive the dynamics. 6) Connecting the stochastic model to the standard deterministic equations of infinite systems. 7) An example of simulation of the limit using Lotka-Volterra equations. 8) Working out the proper stochastic-deterministic connection, using and arguing for the system size expansion. 9) Further comments on the modeling of interaction rates. 10) Outlining the arguments for models of eco-evolutionary dynamics with the generalization of evolutionary game theory. 11)
Sketching and playing with parameter estimation on reaction networks using MCMC. 12) A list of intentions of further improvements. I also include two appendices with the derivation of the master equation and the MCMC algorithm through my own lenses.

\section{Bayesian background}

Suppose we perform an experiment of tossing a coin. Our question of interest is: what is the probability of it landing heads? The simplest widespread answer is $1/2$. This is a parsimonious answer, guided by a principle of indifference, assuming an intuitive understanding of probability as chances or odds of occurrence. We characterize the experiment as having two possible outcomes (heads and tails) and, in the absence of further information, the outcome is symmetrical. Hence, $1/2$. But, first of all, what does it mean for an event to have a probability of $1/2$? Is it a property of the system, of the coin toss? For example, we should expect this random behavior from the coin, intrinsically? The Bayesian framework argues that this is not the case. That probability belongs to whoever is asking, it is the logical evaluation of statements, a numerical assessment of the degree at which a statement is believed to be true, by a model (or "point of view").\cite{Gregory2005BayesianSupport, JaynesProbabilityScience} We are the ones seeing chances; nature just happens. The statement "the outcome of the coin toss is heads" has a truth value given by its probability. The probability of a statement is generated by a model of reality that feeds information to the statement's evaluation. The model for answering $1/2$ a priori is an argument of symmetry between two states. But suppose one tosses the coin and produces several rounds of outcomes, observing the pattern, or the frequencies of outcomes. One could arrive at the following experimental frequencies: $49\%$ of heads, $49\%$ of tails, and $2\%$ of the coin landing by its side. With this information in hand, we can inform our model and update it with a new possible outcome and differing observed chances of occurrences. This would be a frequentist model for evaluating the statement "the outcome of the coin toss is heads (or tails or sides)". Note that the experiment of the coin remains the same, nature remains the same. The frequentist model is more accurate and informed than the a priori model. But there is a third and much better model: suppose one knows the (sufficiently) exact initial state and the laws of motion underlying the coin toss, then one could model the actual dynamics of the coin, predicting its exact outcome. This model would evaluate the statements as having a probability of $0$ or $1$ depending on the toss. This is a perfect information model, able to remove all uncertainty about the experiment and reduce the probability to the binary: true or false. It reflects the experiment in a much more accurate way (with minimum entropy, while the first model is of maximum entropy on two unconstrained states). While these models are more and more accurate on reflecting the outcomes, from a Bayesian perspective they are equally right on making the most out of the available data. In doing science, our experiments yield data that almost always is incomplete to evaluate the statements of interest. For example we toss coins in "randomized" manners, or we collect data with limited precision and scope. This requires us to work with probabilities instead of true/false values.

Acting on sets that are collections of events/statements, probability can be shown to hold two key rules. Sets can be united, $A+B$, and can intersect $A,B$. The rules of probability determine how it transforms over the set operations. The sum rule:
\begin{equation}
   P(A+B) = P(A)+P(B)-P(A,B).
\end{equation}
And the product rule:
\begin{equation}
    P(A,B) = P(A|B)P(B) = P(B|A)P(A).
\end{equation}
The set operation $A|B$ means $A$ given $B$, a conditional, depending on a logical relation between $A$ and $B$ (it means the intersection $A,B$ with the event space given by $B$). Other meaningful relation is the negation probability $P(\overline{A})=1-P(A)$, with $A,\overline{A}=\emptyset$ and $A+\overline{A}=U$ ($U$ is the whole event space). Generalizing this to more than two sets, we draw a partition of the event space given by the sets $E_i$ (for a partition, $E_i,E_j=\emptyset$ for $i\neq j$ and $\sum_i E_i=U$), then the law of total probability shows how to condition any event on the partition:
\begin{equation}
    P(A) = P((\sum_iE_i),A) = P(\sum_iE_i,A) = \sum_i P(E_i,A) = \sum_i P(A|E_i)P(E_i).
\end{equation}
These simple results are the central components of the theoretical frameworks of stochastic calculus and statistics. Moreover, they present a framework for understanding scientific modeling in general. The Bayes equation of probability update follows directly as:

\begin{equation}
    P(A|B) = \frac{P(B|A)P(A)}{P(B)}.
\end{equation}

For example, the equation summarizing the effect of modeling in science is $P(H|D)\propto P(D|H)$. It is the transformation of the assessment of data given hypothesized models into the assessment of hypothesized models given data. Technically, it informs the posterior probability of a hypothesis with the construction of a data likelihood under the hypothesis.

\section{Reaction networks}

I use reaction networks as a theoretical framework and as a design heuristic for models of population dynamics.\cite{Schnoerr2017ApproximationReview, Feinberg2019FoundationsTheory} The elements of populations are considered to have a default independence between themselves, their interactions and all that is able to change states can only happen through well defined reaction processes. The dynamical setting of the populations, determining the changes in the state vector of population sizes ($\boldsymbol{n}$), is given by a uniform probability landscape that is perturbed by the interactions. Interactions are defined at the individual level, reflecting the probabilities associated with their underlying mechanisms. These individual-level processes combine to shape a dynamical landscape at the population level. Thus, we have mechanisms of individual interactions providing chances for removal and addition of elements as shifts in the state-space of population sizes that, in turn, provide together the evolution of the population as a whole.

Populations: collections of elements that are identical on how they interact with the network. For example, a population that is in state $n=n^*$ is composed of $n^*$ elements. The state of a
network is the vector of all population states, $\boldsymbol{n}$.

Given $N$ populations $X_i$ with numbers of elements $n_i$, the reactions are represented as:
\begin{equation}
    \sum_{i=1}^Ns_{ir}X_i\xrightarrow{k_r} \sum_{i=1}^Ns^{\prime}_{ir}X_i, \quad r=1,2...R.
\end{equation}
The $s_{ir}$ and $s^{\prime}_{ir}$ are stoichiometric coefficients, respectively the amount of $X_i$ needed for the interaction and the amount of $X_i$ in the outcome of the interaction. The $k_r$ is a rate parameter that can be further modelled as state-dependent interaction mechanism. It drives the rate at which the interaction occurs.

If we have $X_1 \xrightarrow{\omega_1}2X_1$, then at a rate $\omega_1$ any single element $X_1$ can duplicate, representing a type of self-replication. In this case, we need $s_{1r}=1$ element $X_1$ and the outcome is $s^{\prime}_{1r}=2$ elements. Note that the relevant dynamical effect of the interaction is to cause a shift in the state-space of population sizes.

The definition of a model in terms of reaction networks highlights some important theoretical features of the dynamics: 1) population sizes are the first-principles object of interest for the dynamical analysis. 2) All relevant dynamical events are relational, meaning that what matters is how elements relate, not what they are, so elements and populations are general abstract concepts. 3) The population-level trajectories are ultimately governed only by well-defined individual-level local interactions. 4) The rules determining the consequences of interactions are not physical trajectories or context-dependent encounters, they are only probability structures, with uniformly distributed microstates outside interaction events (at first, we know nothing further about the microstates of the systems).

As an example, a possible predator-prey network can be the following:
\begin{gather}
    X_1 \xrightarrow{\omega} 2X_1 \nonumber\\
    X_1+X_2 \xrightarrow{\beta} (1+\delta) X_2 \nonumber\\
    X_2 \xrightarrow{\mu} \emptyset,
     \label{lotkavolterra}
\end{gather}
The prey $X_1$ can replicate itself at a rate $\omega$, then a prey can be encountered and consumed by a predator $X_2$ at a rate $\beta$, resulting in $\delta$ new predators in the population, then a predator may die and be removed at a rate $\mu$.

\section{Master equation}

If we consider a system of populations with interactions as defined through a reaction network, assuming a continuous passage of time, we have the shifts in state-space given by a Markov jump process. We saw that the shifts are discrete, changing the state $\boldsymbol{n}$ by integer jumps on components (given by stoichiometric coefficients), and the shifts are Markovian because rates are defined to be present state-dependent. Just an observation: it is impossible for nature to be non-Markovian. When models are non-Markovian, it is because we are using a "memory" tool as a proxy to a lack of information about the present state.

The time-evolution of probabilities of the states of the model can be derived using just the probability rules previously defined, and it is given by a master equation of stochastic evolution for Markov jump processes (Appendix 1). Given the transition rates $W_r$ for each reaction, we have:
\begin{equation}
    \frac{d \Pi_{\boldsymbol{n}}(t)}{d t} = \sum_{r}\left(W_{\boldsymbol{n}-S_r^T,\boldsymbol{n}}^r\Pi_{\boldsymbol{n}-S_r^T}(t)-W_{\boldsymbol{n},\boldsymbol{n}+S_r^T}^r\Pi_{\boldsymbol{n}}(t)\right).
\end{equation}
Here, $S_r^T$ is the vector representing the shifts in state-space, with $S_{ir}=(s^{\prime}_{ir}-s_{ir})$. Note that the master equation is additive on the different reactions and it measures the balance between possible jumps into the focal state and possible jumps out of the focal state.

Written as a state operation, the master equation is:

\begin{equation}
    \frac{d \Pi_{\boldsymbol{n}}(t)}{d t}= \Big[\sum_{r=1}^R\Big(\prod_{i=1}^NE_i^{-S_{ir}}-1\Big)W_r(\boldsymbol{n})\Big]\Pi_{\boldsymbol{n}}(t).
\end{equation}
This is in terms of the step operator, that shifts specific components of vectors:
\begin{equation}
    E_i^{S_{ir}}g(\boldsymbol{n}) = g(n_1,...,n_i+S_{ir},...,n_N).
    \label{stepOperator}
\end{equation}
In this setting, any interaction can occur at a given time, causing a jump in state-space, with exponentially distributed waiting times between interactions. The interaction rate $W_r$ weights the frequency of occurrence of the different interactions. Note that the master equation is written as a set of equations, one per state, which reads as the time-evolution of probability densities for fixed states.

\section{Transition rates}

In order to fully characterize the dynamics of interactions, we must determine how reactions $r$ translate into the transition rates $W_r$ governing the jumps between states. For that, let's further specify the condition for interactions to occur. The major assumption is that, outside of interactions, elements are independent and uniformly distributed. Then, the interest is to separate between what is a direct consequence of the dynamics from what is context-dependent and holds the mechanisms underlying the interaction.

We define then an effective volume $\omega_r$ of a reaction $r$ meaning the volume inside which elements are considered to be in contact with each other for reaction $r$. If there is a set of all elements required by $r$ inside this volume, then the local interaction mechanisms can act to give rise to the reaction itself. The chances of relevant groups of elements to be inside $\omega_r$ comprises the dynamical portion of the transition $r$ while the contact per se, given the confluence of elements, comprises the contextual portion of the transition.

The transition rates $W_r$ are global rates measuring the overall propensity of the system to output the reaction $r$, therefore they count the activity in every cell of volume $\omega_r$, and there are $\Omega/\omega_r$ of such cells. Now, since the elements are all independent, we may consider a given combination of $s_r=\sum_i s_{ir}$ elements necessary for the reaction and then sum for every one of the $l$ possible combinations. Then, consider the statements $F_{lr}$=\{A portion of the system, with volume $\omega_r$, holds inside it all $s_r$ elements of the $l$-th combination required to activate reaction $r$\}, and $K_{lr}$=\{The interaction mechanisms of reaction $r$, for the $l$-th combination of elements, are acting on an active site during an interval $dt$ and are allowing the interaction to occur\}. Then, we have:
\begin{equation}
    W_rdt = n_{cells}\sum_lP(F_{lr},K_{lr})=\frac{\Omega}{\omega_r}\sum_lP(K_{lr}|F_{lr})P(F_{lr})
\end{equation}
All quantities are independent of the particular combination of elements and the number of combinations is the product of the $N$ binomials of $s_{ir}$ given $n_i$:
\begin{equation}
    W_rdt = \frac{\Omega}{\omega_r}\sum_lP(K_{lr}|F_{lr})P(F_{lr})=\frac{\Omega}{\omega_r}P(K_r|F_r)P(F_r)\prod_i\binom{n_i}{s_{ir}}
\end{equation}

Note that $K_r$ is a local statement, concerning only a given active site and a given set of $s_r$ elements. We define $P(K_r|F_r)=k^0_rdt$. This means that, at a local rate $k^0_r$, the interaction mechanisms result in the occurrence of a jump. Assuming $F_r$ to remain true for a period $dt$, then the probability of a reaction to occur given an encounter is $k^0_rdt$.

The chances of a single element to be inside a portion $\omega_r$ of a system of size $\Omega$ are $\omega_r/\Omega$. Given a set of $s_r$ independent elements, the joint chances are $P(F_r)=(\omega_r/\Omega)^{s_r}$. We have then the result:
\begin{equation}
W_r(\boldsymbol{n})=k^0_r\frac{\omega_r^{s_r-1}}{\Omega^{s_r-1}} \prod_i\frac{n_i!}{(n_i-s_{ir})!s_{ir}!}.
\end{equation}

Side note: Van Kampen (1991) and Lacroix (2021)\cite{N.G.VanKampenStochasticChemistry, Lacroix2021DichotomyLaw} seem to be solving another problem in their cut of this issue. In their construction, they consider that, for a given site to be active for reaction $r$, all other components must be outside of it, thus including a term
\begin{equation}
    \Big(1-\frac{\omega_r}{\Omega}\Big)^{N-s_r},
\end{equation}
with $N=\sum_in_i$. They argue that the term will vanish for $\omega_r<<\Omega$, but it doesn't, and it will depend on the densities of populations even in the limit of infinite system (it becomes $e^{-\eta\omega_r}$ in the deterministic limit, where $\eta$ is the total density of populations; it is a constraint penalizing interactions for being broadly ranged or a system that is too crowded, because they make it harder for other elements to be outside $\omega_r$ as required). Our argument, however, is that it is indeed a different problem. Should we have to demand all other components to be outside the $\omega_r$? This is in contrast with the fact that interactions are benefited by the increase of densities. And since all components are independent, the presence of exceeding components should be at most irrelevant, but also possibly beneficial for interactions if these components can be used as reactants. Also, in this case there is no need to assume $\omega_r<<\Omega$, only that $\omega_r<\Omega$.

Continuing. We just derived the transition rates $W_r$ in terms of the state-vector, the interactions' dynamical parameters (stoichiometric coefficients and the effective volume or range of interaction), and the system size. The result can be called a stochastic law of mass action. Apart from the reaction mechanisms encoded through $k^0_r$, the result is general and presents the dynamical rule of motion followed as a consequence of interactions, with the elements of populations navigating independently and according to uniformly distributed microstates apart from the reactions. Now, the transition rates can be rearranged as:
\begin{equation}
    W_r(\boldsymbol{n}) = \Big(\frac{k_r^0\omega_r^{s_r-1}}{\prod_is_{ir}!}\Big)\Omega\prod_i\frac{n_i!}{(n_i-s_{ir})!\Omega^{s_{ir}}}.
    \label{law}
\end{equation}
We recognize the quantity in parenthesis as the standard reaction rate (or interaction rate, that goes over the arrow on the representation of the reaction):
\begin{equation}
    k_r = k_r^0\frac{\omega_r^{s_r-1}}{\prod_is_{ir}!}.
    \label{interactionrate}
\end{equation}
This absorbs the dependence on $\omega_r$, which conceals its role in the dynamics. Note that the exponent of $\omega_r$ is $s_r-1$, so for interactions of order 1 ($s_r=1$) there is no dependence, we have $W_r=k^0_rn_i$ (if the reaction needs a $X_i$ element). This makes sense, because in this case $\omega_r$ has no meaning; these interactions are dynamically spontaneous, happening within individuals, and $k^0_r$ is the intrinsic rate of reaction for each element of $X_i$. For $s_r=0$, then $W_r=k^0_r\Omega/\omega_r$. This is because every one of the $\Omega/\omega_r$ cells of volume $\omega_r$ is a "source" of the reaction with propensity $k^0_r$. If the source is homogeneous in the whole space, we should have $k^0_r \propto \omega_r$, then $W_r\propto\Omega$, and the chances of the reaction are simply proportional to the system size.

\section{The stochastic-deterministic connection}

The deterministic limit is the thermodynamics of the infinite system, where the stochastic dynamics is replaced by precise trajectories of population densities. If we infinitely replicate the system and ask about its finite properties, we get the deterministic equations. The change of state-vector from the stochastic counts of elements to the deterministic densities of populations is done through a specific limit:
\begin{equation}
    \eta_i = \lim_{(n_i,\Omega)^*\to\infty}\frac{n_i}{\Omega}
    \label{limit}
\end{equation}
Valid for every population. The pair $(n_i,\Omega)$ is not going to infinity in any way, this is the same system of size $\Omega$ being infinitely extended, keeping its local configuration the same, also like considering an ensemble of systems. In this case, their ratio is finite and equal to the population density of the infinite system. Then we talk about intensive (local) quantities. The analogous quantity of the transition rates, that ultimately changes densities as a consequence of local reactions occurring, is given by:
\begin{equation}
    W^d_r = \lim_{(n_i,\Omega)^*\to\infty} \frac{W_r}{\Omega}=k_r\prod_i\eta_i^{s_{ir}}.
\end{equation}
This becomes the law for the rate at which each reaction happens in the infinite system. The link between the rate of reactions and the change in densities is the size of the jump in state-space $S_{ir}=s'_{ir}-s_{ir}$. Thus, the deterministic laws of motion become:
\begin{equation}
    \frac{d\eta_i}{dt} = \sum_rk_rS_{ir}\prod_j \eta_j^{s_{jr}}.
    \label{detlaw}
\end{equation}
This is the deterministic and usual law of mass action. In the infinite system it should, in this form with the black-box on $k_r$, be easily derived just by invoking independence between elements (the term $\prod_j\eta_j^{s_{jr}}$ is simply counting the presence of all required elements independently and connected by AND logical operations, without any regard for reposition). Again, this is a general law and all context-dependent dynamics representing interaction mechanisms should be present in models of $k_r(\boldsymbol{\eta})$. This connection marks an important disambiguation of the deterministic law of motion. Here, equation (\ref{detlaw}) is a consequence of a reaction network defined locally in terms of relations between actual elements, reactions shaping the evolution of states. If we define models directly through the deterministic law, we lose the structure of interactions and the model becomes potentially ambiguous. Let's go back to the earlier example of predator-prey interactions. The deterministic system of the network in (\ref{lotkavolterra}) is:
\begin{equation*}
    \Dot{\eta_1} = \omega\eta_1-\beta\eta_1\eta_2
\end{equation*}
\begin{equation}
    \Dot{\eta_2} = \delta\beta\eta_1\eta_2-\mu\eta_2.
\end{equation}
Then, note the subtle structure of the predator growth rate $\delta\beta$, highlighting the fact that there are two sources at play: the prey-consuming rate and the conversion rate between prey and predator. For a more intricate example, take a logistic growth with a standard Allee effect, as modeled in terms of a deterministic equation:
\begin{equation}
    \Dot{\eta} = r\eta\Big(1-\frac{\eta}{K}\Big)\Big(\frac{\eta}{A}-1\Big).
\end{equation}
A general idea of an Allee effect should not necessarily require any structure breaking the assumptions of independence and homogeneous distribution of elements. But then what is the combination of interactions and models of $k_r(\boldsymbol{\eta})$ giving rise to this motion? There are various possibilities, each one meaning something different locally and being backed by a different stochastic law.

\section{Simulation of the Limit}

To fully appreciate this connection, we consider a simple example where we show a stochastic trajectory, simulated from the stochastic simulation algorithm (SSA),\cite{Gillespie1976AReactions} increasingly approaching the deterministic trajectory of the same system as the system size increases. For that, we'll use the simple predator-prey system shown before
\begin{gather}
    X_1 \xrightarrow{\alpha} 2X_1 \nonumber\\
    X_1+X_2 \xrightarrow{\beta} (1+\delta) X_2 \nonumber\\
    X_2 \xrightarrow{\gamma} \emptyset.
\end{gather}
This system exhibits well-known oscillatory trajectories with differing phases (preys go up, then predators go up, then preys go down, then predators go down). The deterministic differential equations are given in (\ref{lotkavolterra}), and the full-length master equation is, with implicit dependence on time
\begin{equation*}
    \frac{d\Pi(n_1,n_2)}{d t}= \alpha\Big( (n_1-1)\Pi(n_1-1,n_2)-n_1\Pi(n_1,n_2) \Big)
\end{equation*}
\begin{equation*}
    +\frac{\beta}{\Omega}\Big( (n_1+1)(n_2-\delta)\Pi(n_1+1,n_2-\delta)-n_1n_2\Pi(n_1,n_2) \Big)
\end{equation*}
\begin{equation}
    +\gamma\Big( (n_2+1)\Pi(n_1,n_2+1)-n_2\Pi(n_1,n_2) \Big).
\end{equation}
Each line shows the contribution from one reaction, with the first term being the jump into the state $\boldsymbol{n}=(n_1,n_2)^T$, and the second term being the jump out of that state.

\begin{figure}[ht]
    \centering
    \includegraphics[width=14cm]{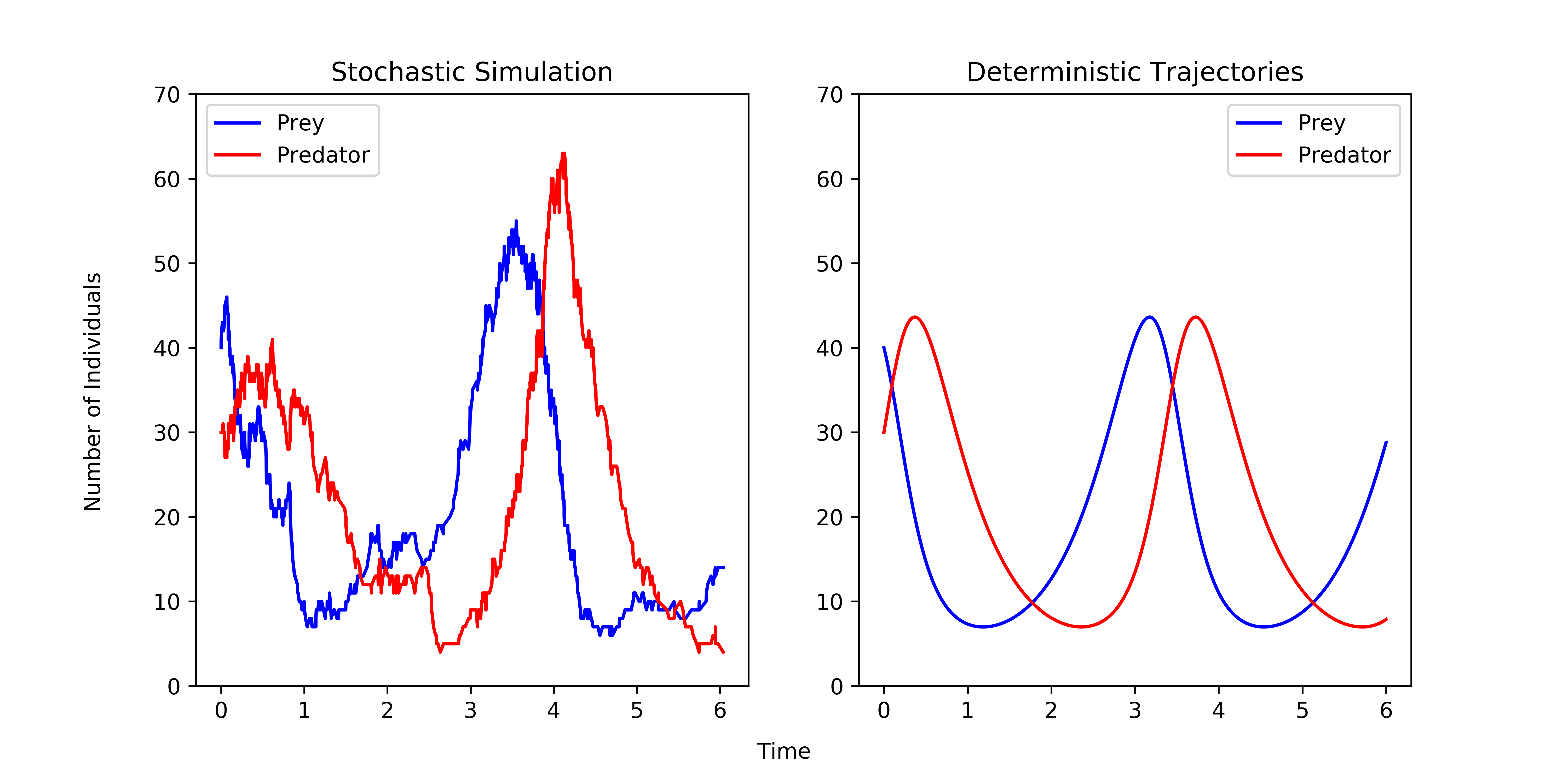}
    \caption{Predator-Prey reaction network. Left: a sample of the master equation from the SSA. Right: a numeric solution of the deterministic equation. The system size is $\Omega=1$, so both scales coincide. Initial values are $n_1=\eta_1=40$ and $n_2=\eta_2=30$. With arbitrary time-scale, we have $\alpha=\gamma=2$, $\beta=0.1$, and $\delta=1$.} 
    \label{pred-prey-det-stoch}
\end{figure}

\begin{figure}[ht]
    \centering
    \includegraphics[width=14cm]{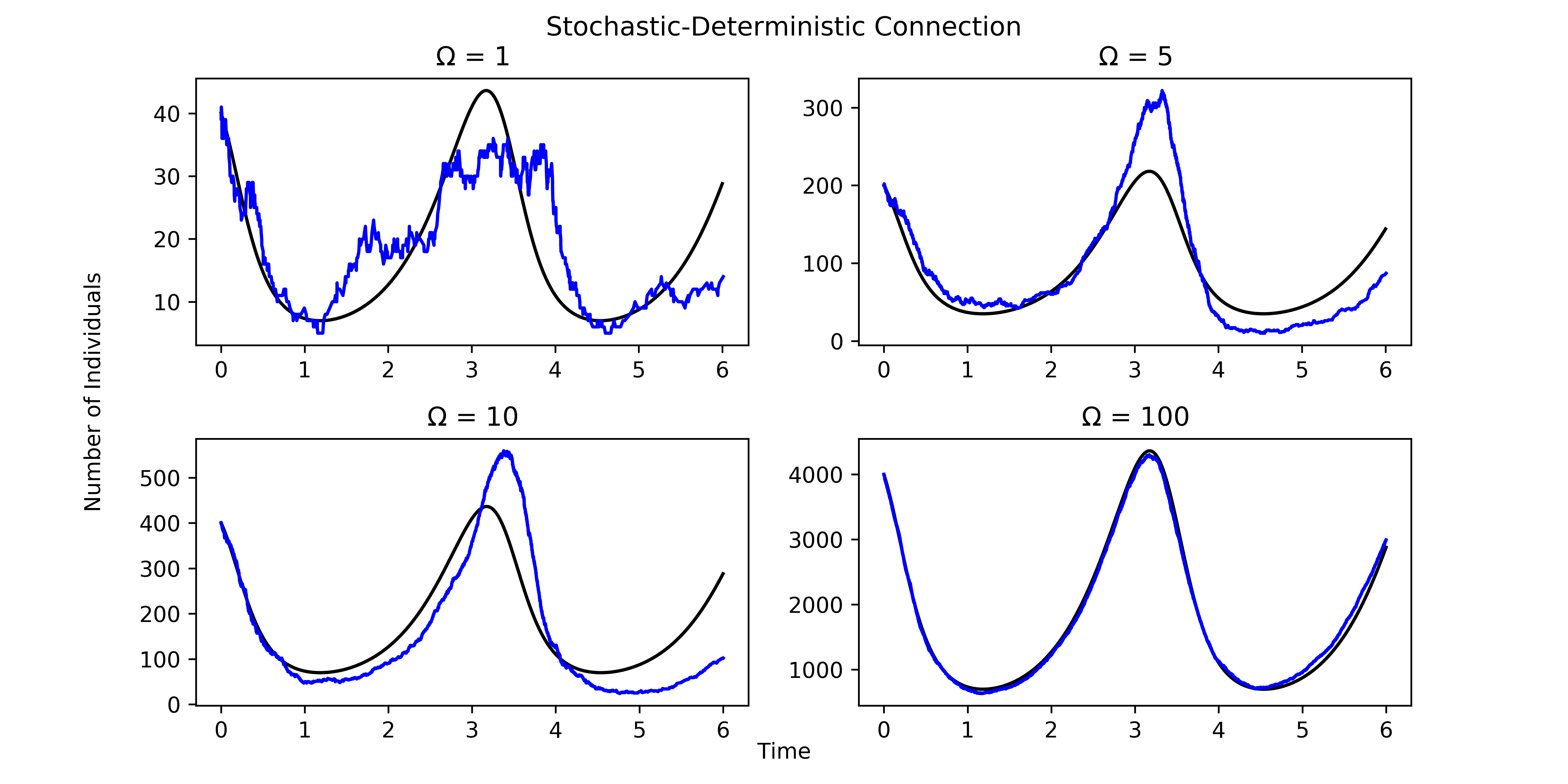}
    \caption{Number of preys compared for the four system sizes $\Omega=1,5,10,100$. The deterministic solution ($\Omega\eta_1$) is shown in black and the stochastic samples from the SSA ($n_1$) are shown in blue. Initial values of concentrations are kept at $\eta_1=40$ and $\eta_2=30$, and the initial number of individuals are scaled accordingly, $n_1=40\Omega$ and $n_2=30\Omega$. With arbitrary time-scale, we have $\alpha=\gamma=2$, $\beta=0.1$, and $\delta=1$.} 
    \label{connection}
\end{figure}

In figure \ref{pred-prey-det-stoch}, we show a sample of the stochastic system together with the trajectories of the correspondent deterministic system, for a system size of $\Omega=1$. We see the contrast between both dynamics with such a low system size. Figure \ref{connection} shows the connection between them as the system gets bigger, for the trajectory of preys. We maintain the same initial deterministic concentration and other parameter values through all four scenarios. Note how, at size $\Omega=100$, both dynamics already are hardly distinguishable.

One final aspect to note that is related to this connection is the evolution of the mean number of elements of a species, also given by the master equation. The mean is defined as
\begin{equation}
    \langle n_i \rangle(t) = \sum_{\boldsymbol{n}} n_i \Pi_{\boldsymbol{n}}(t).
\end{equation}
where the sum extends over all possible combination of states $\boldsymbol{n}$. In the same way, the mean extends to any function of $\boldsymbol{n}$,
\begin{equation}
    \langle g(\boldsymbol{n}) \rangle = \sum_{\boldsymbol{n}} g(\boldsymbol{n}) \Pi_{\boldsymbol{n}}(t).
\end{equation}
To find the equation for the mean, we produce this definition over the master equation. We multiply the whole equation by $n_i$ and sum over all possible states:
\begin{equation*}
    \frac{d \sum_{\boldsymbol{n}} n_i\Pi_{\boldsymbol{n}}}{d t}= \sum_{r=1}^R\Big(\prod_{i=1}^NE_i^{-S_{ir}}\sum_{\boldsymbol{n}}(n_i+S_{ir})W_r(\boldsymbol{n})\Pi_{\boldsymbol{n}}-\sum_{\boldsymbol{n}}n_iW_r(\boldsymbol{n})\Pi_{\boldsymbol{n}}\Big),
\end{equation*}
\begin{equation}
    \frac{d \langle n_i \rangle}{d t}= \sum_{r=1}^R\Big(\prod_{i=1}^NE_i^{-S_{ir}}\langle(n_i+S_{ir})W_r(\boldsymbol{n})\rangle-\langle n_iW_r(\boldsymbol{n})\rangle\Big).
\end{equation}
note how $n_i$ transforms into $n_i+S_{ir}$ when passing inside the step operation that discounts the shift by $S_{ir}$. The step operator also doesn't shift anything in means, because they don't depend on the system state $\boldsymbol{n}$, so it just vanishes. Then the $n_i$ portion of the first term cancels out with the second term, and we end up with
\begin{equation}
    \frac{d \langle n_i \rangle}{d t}= \sum_{r=1}^RS_{ir}\langle W_r(\boldsymbol{n})\rangle.
\end{equation}
This equation has the same form as the deterministic system from equation (\ref{detlaw}), and with it we can see how $\langle n_i \rangle/\Omega$ is equal to $\eta_i$ in the limit from the point of view of the master equation. We also see that, if all reactions are of type $\emptyset \xrightarrow{}?$ and $X_1 \xrightarrow{}?$ (both zero and first order reactions, the ones that give linear transition rates), the equation for the average is exactly the same as the deterministic equation for $\eta_i$. We could be tempted to say that, at least in these cases, they are the same even without the limit; even for small systems. But this is wrong! The equality $\eta_i=\langle n_i \rangle/\Omega$ never holds without the limit, it doesn't even make sense. At the level of $\eta_i$, we don't have a finite value of $\Omega$ (it doesn't exist as a parameter). And if we define $\eta$ without the limit, we are approximating with error of order $1/\Omega$ (an error we don't even have access to if we don't have access to $n_i$ or $\Omega$). But it is remarkable that the equations of motion have the same form for linear systems, even for small systems, and are equal up to a constant factor. It means that being small doesn't alter the shape of the average dynamics of linear systems, and the same doesn't occur for nonlinear systems.

But, how can we really interpret mean motion? It is the motion we would expect uncertainties to be placed around, an unbiased estimator, while $\eta_i$ is what we would expect to see in very large systems. But, then, for large systems, $\eta_i$ also is where we would place uncertainties around (the two kinds of motion converge as the system gets ever larger and uncertainties shrink).

\section{System-size expansion}

We still need to offer a legitimate path from the microscopic system to the macroscopic motion of the infinite system. The ideal is to have a systematic expansion in terms of a small parameter asymptotically approaching the deterministic law as the system grows. Since we already have a parameter that represents the size of the system and grows to infinity, the volume $\Omega$, it is natural to expand on a function of $\Omega$. Van Kampen's ansatz\cite{N.G.VanKampenStochasticChemistry} is a means of incorporating a law of large numbers in order to rewrite the system in terms of a noise that diminishes as the system grows. Then, the scale of the noise in terms of the system size is a good candidate of small parameter. The ansatz is a sort of generalization of the limit in (\ref{limit}), decoupling the element count into a deterministic trajectory and the noise around it. As in the law of large numbers, with the counting modulated by independent reactions, the variance of $n_i$ should scale as $n_i$, making the noise (standard deviation) scale as the square root of $n_i$'s scale. Thus, we make:
\begin{equation}
    n_i = \Omega \eta_i + \sqrt{\Omega}\xi_i.
\end{equation}
The $\xi_i$ is the noise around $\eta_i$. This ansatz will be further justified later when we verify its self-consistency. Given the change between $n_i$ and $\xi_i$, the expansion consists of calculating the master equation in terms of successive orders of $\Omega^{-1/2}$.

The expansion follows like this:
\begin{equation}
    g(\boldsymbol{n}/\Omega) = g(\boldsymbol{\eta}+\Omega^{-1/2}\boldsymbol{\xi})=g(\boldsymbol{\eta})+\Omega^{-1/2}\sum_j\xi_j\frac{\partial g}{\partial \eta_j}(\boldsymbol{\eta})+\mathcal{O}(\Omega^{-1}).
\end{equation}

The change of distributions from $\boldsymbol{n}$ to $\boldsymbol{\xi}$, $\Pi_{\boldsymbol{n}}\to\Pi^*_{\boldsymbol{\xi}}$, is:
\begin{equation}
    \frac{d \Pi_{\boldsymbol{n}}}{d t} = \frac{d\Pi^*_{\boldsymbol{\xi}}}{dt}+\sum_j\frac{d\xi_j}{dt}\frac{d\Pi^*_{\boldsymbol{\xi}}}{dt}=\frac{d\Pi^*_{\boldsymbol{\xi}}}{dt}-\Omega^{1/2}\sum_j\frac{d\eta_j}{dt}\frac{d\Pi^*_{\boldsymbol{\xi}}}{dt}.
\end{equation}
The step operator $E^{S_{ir}}$ shifts states by $\Omega^{-1/2}S_{ir}$ instead of $S_{ir}$. Finally, we must also rewrite the transition rates in terms of $\boldsymbol{n}/\Omega$, which is always possible to do in the context of reaction networks (this is important):
\begin{equation}
    W_r(\boldsymbol{n})=\Omega\sum_l\Omega^{-l}W_r^{(l)}(\boldsymbol{n}/\Omega).
\end{equation}
This may not be possible when further modeling $k^0_r$ in terms of $\boldsymbol{n}$, but it should make more sense to model directly in terms of $\boldsymbol{\eta}$, because this will be already interpreted in the network in terms of a mean-field propensity.

Putting all together in the master equation, we arrive at the successive orders of the system size expansion. There is a term of order $\Omega^{1/2}$ that should vanish if we want the expansion to make sense. Indeed, this term vanishes precisely if the deterministic law (\ref{detlaw}) is valid, so this is the self-consistency of the ansatz. The first order of approximation is called linear noise approximation (LNA) and it is a sort of central limit theorem, approximating the first order of noise to a time-dependent normal distribution. The equation has the form of a Fokker-Planck equation:
\begin{equation}
    \frac{d \Pi^*_{\boldsymbol{\xi}}(t)}{d t}=\sum_r\Big( -\sum_{i,j}S_{ir}\xi_j\frac{\partial W^{(0)}_r(\boldsymbol{\eta})}{\partial \eta_j}\frac{\partial}{\partial\xi_i}+\frac{1}{2}\sum_{i,j}S_{ir}S_{jr}W^{(0)}_r(\boldsymbol{\eta})\frac{\partial}{\partial \xi_i}\frac{\partial}{\partial \xi_j} \Big)\Pi^*_{\boldsymbol{\xi}}(t)+\mathcal{O}(\Omega^{-1/2}).
    \label{LNAeq}
\end{equation}
It can be solved by finding the first two moments of the distribution. For deterministic initial condition, the mean is always zero, and we only have to calculate the variance. This is done by solving a Lyapunov equation on the covariance matrix using only the solution to the deterministic equations. The equation is the matrix form of:
\begin{equation}
    \frac{d\langle \xi_i\xi_j \rangle}{dt}=\sum_r \Big(
    \sum_j \frac{\partial W^{(0)}_r(\boldsymbol{\eta})}{\partial \eta_j} \big[S_{kr}\langle \xi_k\xi_j \rangle +  S_{lr}\langle \xi_l\xi_j \rangle \big]+
    \frac{1}{2}S_{kr}S_{lr}W^{(0)}_r(\boldsymbol{\eta}) \Big).
    \label{LNAvar}
\end{equation}

Defined in terms of the deterministic system solutions, through the Jacobian and a diffusion matrix:
\begin{equation}
    \Sigma_{ij}=\langle \xi_i\xi_j \rangle
\end{equation}
\begin{equation}
    D_{ij} = \sum_r S_{jr}\frac{d\eta_i}{dt} = \sum_r S_{ir}S_{jr}W^d_r(\boldsymbol{\eta})
\end{equation}
\begin{equation}
    J_{ij} = \frac{\partial}{\partial \eta_j} \frac{d\eta_i}{dt} = \frac{\partial}{\partial \eta_j} \sum_r S_{ir}W^d_r(\boldsymbol{\eta}),
\end{equation}
\begin{equation}
    \frac{d\Sigma}{dt} = J\Sigma+\Sigma J^T + \Omega D.
\end{equation}

Note: There are other ways to approximate the master equation, but they are not systematic. The Kramers-Moyal expansion is a method that directly applies a Taylor expansion to the equation. The Pawula theorem states that, in order for the approximated solution to be a proper probability distribution, only the first two expansion terms can be considered. Also, in order to apply the expansion, the count-number variable $\boldsymbol{n}$ is transformed into a continuous variable. So the expansion really becomes just a coarser continuous picture that follows the same propensities as the master equation. The resulting equation for a continuous state variable is a usual Fokker-Planck equation (different from the LNA solution), that has the two expansion terms identified as a drift and a diffusion contributions. This equation, in turn, holds an equivalency to the stochastic differential equation picture represented by an associated Langevin equation. The Fokker-Planck also connects to a path integral formulation. Interestingly, we can instead apply the SSE to this continuous approximation of the master equation and also obtain the LNA, with the drift contribution giving rise to the deterministic trajectory and the diffusion contribution giving rise to the noise. That is because the continuous approximation already is implicit in the SSE too. This expansion is also called a diffusion approximation, and it doesn't regard the system size, so it doesn't connect with the deterministic limit and doesn't measure approximation errors.

\section{Discussion on reaction rates and modeling}

Here we further discuss interesting aspects of reaction rates and how we can further model them.

1. Suppose we want to directly compare the rates of a reaction $X \xrightarrow{c}?$ with a reaction $X+Y \xrightarrow{c'}?$. If we want them to have the same mechanism rate, this is not the same as $c=c'$. What we actually want to compare is the $k^0_r$. For example, let's write the condition for a first-order reaction to happen more frequently than a second-order reaction, given that their local mechanisms operate at the same rate ($k^0_r)$:
\begin{equation*}
    k^0_r n_x > k^0_r \frac{\omega_r}{\Omega}\frac{n_xn_y}{2}
\end{equation*}
\begin{equation}
    \omega_r < \frac{2\Omega}{n_y}.
\end{equation}
This means that larger effective volumes of reactions can make it easier for encounters that at first should be rarer. Thinking about it, we see that it should be more common to expect one $X$ element than to expect both a $X$ element and a $Y$ element to encounter. However, the number of possible combinations for the $X+Y$ encounter is larger. In the case that $\omega_r=\Omega$ what happens is clearer. There is a tradeoff between the need of convergence by elements coming inside $\omega_r$ a the same time and the number of possible groups of elements that could independently activate the reaction. Another way to visualize this is to consider the point of view of a single element $X$. By itself, it has a propensity $k^0_r$ to undergo a first-order reaction, but in face of $n_y$ $Y$ elements it would have $n_y$ times that propensity for a second-order reaction provided that the encounter is trivial (for example if $\omega_r=\Omega$). Now, if $\omega_r<<\Omega$, then it is much harder to have matches for the $X+Y$ encounter (whereas the $X$ "encounter" remains trivial), and the second-order reaction becomes rarer.

2. Given that the assumptions of independence outside reactions and uniform distribution hold, the law of mass action as stated should remain valid. However, underlying mechanisms, possibly happening on other temporal or spatial scales, or even characterized in average terms, could in principle shape the dynamics by being ultimately state-dependent. The final form of transition rates (and deterministic equations) would not be read as mass-action, but the dynamical structure of the system would remain the same. The difference is that now $k^0_r=k^0_r(\boldsymbol{n})$. There are many ways to mechanistically justify the modeling of $k^0_r$, but in the end the connection with the population model is probabilistic. We can use a quasi-steady-state approximation to construct a Hill function or Michaelis-Menten function\cite{Santillan2008OnNetworks, GoodwinOscillatoryProcesses, BrianIngalls2014MathematicalIntroduction} for $k^0_r$, in order to justify how the underlying mechanisms actually "underlie" the dynamics. But then, given this validation to the model, what we are telling to the network is that the rate is shaped to assign differential weights to the states, so as to respond in intensity as a function of state. These weights drive the propensities of occurrence of reactions, but are not necessarily tied to the physical collisions. In summary, the point to make is this one: it is not about the physical contact between elements, but how much the consequence of the contact generates propensity, given a state $\boldsymbol{n}$. This line of reasoning opens the case for integration between this framework and games, which means an integration between evolution and ecology.

3. The dynamics of the network is highly dependent on the reactants, i.e. the elements needed for reactions, but not on the products of reactions. Products play the role of just changing the state, but their structure has no part in the jumps. This means that any reaction may generate abstract quantities which are irrelevant to the dynamics because they are never reactants. With this, we can count reactions and also generate payoffs and currencies in reactions.

\section{Population dynamics and EGT}

For this section, I refer to the manuscript (under review): "The eco-evolutionary forces of density-dependent population games". Here, I just outline the main ideas. The aim is to integrate and expand the methods of evolutionary game theory while generalizing the replicator dynamics in terms of actual population densities.\cite{Doebeli2017TowardsTheory, Argasinski2018InteractionDynamics, Krivan2018BeyondGames}

The arguments go as follows. Considering the standard of logistic growth and the implementation of laws of population ecology as discussed by Turchin,\cite{Turchin2001DoesLaws} namely exponential growth and indirect competition for limited resources, I propose a logistic network at the evolutionary scale. But first let's build a connection to the replicator equation, considering only a birth-death network. Thus, a set of populations $\{X_i\}$ evolves as
\begin{equation*}
    X_i \xrightarrow{\omega_i}2X_i
\end{equation*}
\begin{equation}
    X_i \xrightarrow{\mu_i}\emptyset,
    \label{birthdeath}
\end{equation}
yielding the following system of deterministic equations:
\begin{equation}
    \frac{d\eta_i}{dt}=\eta_i(\omega_i-\mu_i)=\eta_iF_i
    \label{densityNomutation}
\end{equation}
where $F_i=(\omega_i-\mu_i)$ is defined in general as a growth function. If we define the proportions as $\rho_i=\eta_i/\sum_j\eta_j$, we can derive the particular replicator equations, that have the general form:
\begin{equation}
    \frac{d\rho_i}{dt} = \rho_i\Big((\omega_i-\mu_i) - (\overline{\omega}-\overline{\mu})\Big)=\rho_i(F_i-\overline{F}),
    \label{replicator}
\end{equation}
with $\overline{\omega}=\sum_i\rho_i\omega_i$ and equivalently for $\overline{\mu}$. Note that the motion of proportions is constrained to the motion of densities, and densities evolve according to the absolute growth while proportions evolve according to relative growth against the average. Also, only in some particular cases the dynamics of densities can be reduced to the dynamics of proportions in a closed form.

Then, we can add the background competition between individuals sharing renewable resources with full capacity limited by use:
\begin{equation}
    X_i+X_j \xrightarrow{\gamma}X_i,
    \label{evonetq}
\end{equation}
where $\gamma$ is a background competition rate. With these reactions, we have an evolution of logistic form. Redefining $f_i=\omega_i-\mu_i$ now as an intrinsic growth rate, this network results in growth rates of the form $F_i=f_i-\gamma\sum_j\eta_j$. In the absence of ecological games (considering the $\omega_i$ and $\mu_i$ as constant parameters) a global equilibrium of densities is achieved only when all intrinsic growth rates $f_i$ are the same, with $(\sum_i\eta_i)_{eq}=f_i/\gamma$. If they differ, the one with higher growth will dominate alone, driving others to extinction. Thus, coexistence between populations is only possible in this case through a frequency-dependent ecological structure, that will dynamically shape $f_i$. Note that the replicator dynamics in terms of proportions stays the same with the inclusion of competition.

The ecological games are then models of the growth and death rates as functions of the state, considering the ecological stage as an average field of interactions where a lifetime (a generation) happens within a single time-step of the evolutionary network.

In a standard game, individuals generate gain and loss payoff instead of being born and dying, so densities remain constant. If the duration of the game is $T$, then the model of birth and death rates in terms of payoffs is given by
\begin{equation*}
    \omega_i(\boldsymbol{\eta},T)=\frac{1}{\eta_iT}\int_0^T \frac{dg_i}{dt}dt
    \label{payfit}
\end{equation*}
\begin{equation}
    \mu_i(\boldsymbol{\eta},T)=\frac{1}{\eta_iT}\int_0^T \frac{dl_i}{dt}dt.
    \label{payfit2}
\end{equation}
The functions $g_i$ and $l_i$ are positive and negative payoffs generated to individuals as a result of ecological interactions. If we calculate these in terms of usual reaction network interactions, for a time-independent average game with stationary payoff generation rates, the model reduces to
\begin{equation*}
    \omega_i=\frac{1}{\eta_i}\sum_r k_r\prod_i\eta_i^{s_{ir}}c^G_{ir},
\end{equation*}
\begin{equation}
    \mu_i=\frac{1}{\eta_i}\sum_r k_r\prod_i\eta_i^{s_{ir}}c^L_{ir}.
\end{equation}
The coefficients $c$ represent the amount of gain and loss payoffs generated to individuals at a given reaction.

A standard form matrix of pairwise interactions is given by:
\begin{align}
\bordermatrix{~ & X_1 & X_2 \cr
                  X_1 & (a^G-a^L) & (b^G-b^L) \cr
                  X_2 & (c^G-c^L) & (d^G-d^L) \cr}.
                   \nonumber
\end{align}
And this translates into a network written as
\begin{equation*}
    2X_1 \xrightarrow{k}2X_1+a^GG_1+a^LL_1
\end{equation*}
\begin{equation*}
    X_1+X_2 \xrightarrow{k}X_1+X_2+b^GG_1+b^LL_1+c^GG_2+c^LL_2
\end{equation*}
\begin{equation}
    2X_2 \xrightarrow{k}2X_2+d^GG_2+d^LL_2.
    \label{ecologicalgame}
\end{equation}
Note that all interaction rates $k$ are made equal and the ecological stage allows propensities for birth and death only from interactions between two individuals. The model of birth-death rates for $X_1$ is then:
\begin{equation*}
    \omega_1 = k(a^G\eta_1+b^G\eta_2)=k(\eta_1+\eta_2)(a^G\rho_1+b^G\rho_2)
\end{equation*}
\begin{equation}
    \mu_1 = k(a^L\eta_1+b^L\eta_2)=k(\eta_1+\eta_2)(a^L\rho_1+b^L\rho_2),
    \label{gamerates}
\end{equation}
and analogously for $X_2$. This model results in the same phase-portrait as a standard game in EGT (differing only in the shaping of the time-scale by the total density). Without evolutionary competition, however, this population has an artificially unstable density growth. Whenever $\omega_i>\mu_i$, densities tend to grow indefinitely, and whenever $\omega_i<\mu_i$, densities tend to decrease indefinitely. For the case of a Hawk-Dove game, $\omega_i>\mu_i$ always for both Hawks and Doves, so densities will rapidly diverge. 

Define the intrinsic growth intensity from ecological interactions as $\phi_i=(\omega_i-\mu_i)/\sum_j\eta_j$, with average, $\overline{\phi}=\sum_i\rho_i\phi_i$, equal to
\begin{equation}
    \overline{\phi}=k\Big(\rho_1\big((a^G-a^L)\rho_1+(b^G-b^L)\rho_2\big)+\rho_2\big((c^G-c^L)\rho_1+(d^G-d^L)\rho_2\big)\Big),
    \label{phidef}
\end{equation}

Now also considering evolutionary competition (the full logistic network), the total density $(\eta_1+\eta_2)$ is at a finite equilibrium when the following condition is met:
\begin{equation}
    \overline{\phi^{eq}}=\gamma.
    \label{gammacondition}
\end{equation}
This is an artificially narrow condition, indicating an inconsistency in the model formulation. The factor $\overline{\phi^{eq}}$ is determined for many games independently of $\gamma$. If $\overline{\phi^{eq}}<\gamma$, the equilibrium total density is zero, and if $\overline{\phi^{eq}}>\gamma$, it still diverges.

The issue is that the ecological interactions should represent a shift from a baseline birth and death that is the exponential growth of the unperturbed populations, modeled in the game with the addition of the reactions
\begin{equation*}
    X_i \xrightarrow{\omega_{0i}} X_i + G_i
\end{equation*}
\begin{equation}
    X_i \xrightarrow{\mu_{0i}} X_i + L_i.
    \label{baselineBD}
\end{equation}
Note: considering this correction without the evolutionary competition is not sufficient to stabilize densities (we can see this by inspection of the equilibrium).

By putting it all together, we arrive at a neat balance equation for the total density of populations at equilibrium in terms of the forces of exponential birth and death, competition for limited resources, and ecological interactions:
\begin{equation}
    \Big(\eta_1+\eta_2\Big)_{eq} = \frac{(\omega_0-\mu_0)}{(\gamma-\overline{\phi^{eq}})}.
    \label{eqDensity}
\end{equation}
It demands that $\overline{\phi^{eq}}<\gamma$ for limited growth, otherwise resource-usage would not be saturated, meaning that resources are effectively unlimited. This is a very informative and simple equation, it poses the total size of populations of an ecological context as a balance between three distinct evolutionary forces.

\begin{figure}[h!]
    \centering
    \includegraphics[scale=0.32]{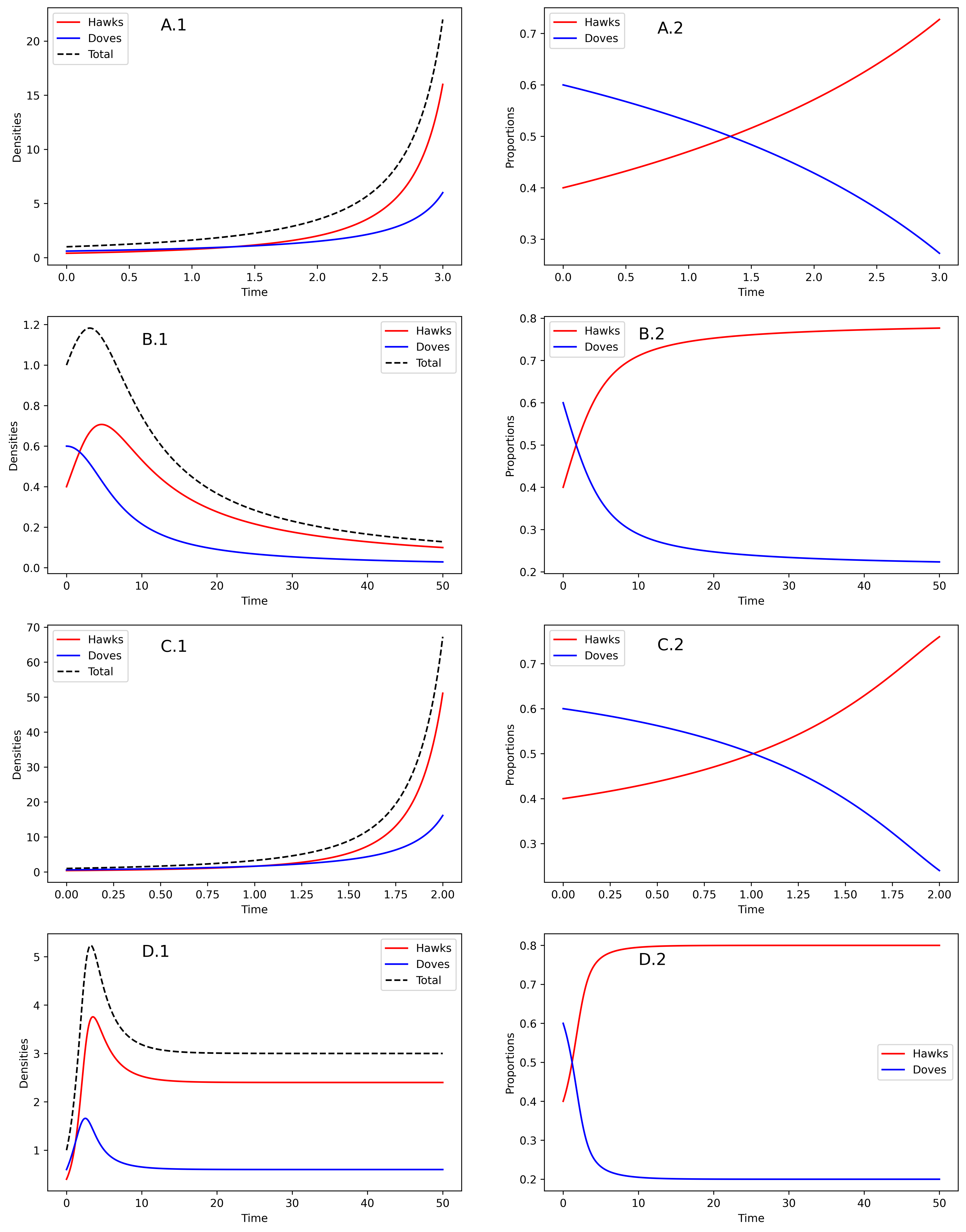}
    \caption{\textbf{Densities and proportions in the Hawk-Dove game}. Left: Densities; Right: Proportions. Initial state for hawks and doves is $(0.4,0.6)$, with densities matching proportions. Equilibrium proportions are the same, but their time-evolution depends on the total density. We have $\overline{\phi^{eq}}=0.1$. \textbf{(A)} Default replicator dynamics. Densities quickly diverge, with unbounded growth, while proportions approach the expected equilibrium. \textbf{(B)} Dynamics with a constant background competition of rate $\gamma=0.3>\overline{\phi^{eq}}$. Now, population size decreases to zero while proportions remain approaching the expected equilibrium values. \textbf{(C)} With a lower competition rate $\gamma=0.05<\overline{\phi^{eq}}$, densities also diverge. \textbf{(D)}  Dynamics with both competition of $\gamma=0.3$ and background birth and death rates for hawks and doves, $(\omega_0-\mu_0)=0.6$. The model reflects a consistent evolutionary setting and the densities are finally stable. Other parameter values: $k=1$, $v=1$, $c=1.25$.}
    \label{HD}
\end{figure}

As an example, consider the Hawk-Dove game given by the interaction matrix
\begin{align}
\label{mhd}
\bordermatrix{~ & H & D \cr
                  H & \frac{(v-c)}{2} & v \cr
                  D &  0 & \frac{v}{2} \cr}
                  \\ \nonumber
\end{align}
with $c>v$. For this simple game we have
\begin{equation}
    \overline{\phi^{eq}}=\frac{kv(c-v)}{2c}.
\end{equation}
Then, the stabilized densities are
\begin{equation}
    (\eta_H+\eta_D)_{eq}=\frac{2c(\omega_0-\mu_0)}{2c\gamma-kv(c-v)}.
    \label{stable}
\end{equation}
See in figure (\ref{HD}) the illustration of this discussion.

Now, looking back at how the framework is built, we see how we are able to immediately derive stochastic forms of eco-evolutionary dynamics, also with a much more general definition of games, including diverse and state-dependent interaction rates, with interaction groups of any size. We can also couple adaptive dynamics for evolution of traits to this framework\cite{Brannstrom2013TheDynamics}, allowing the emergence of mutants at the equilibrium of densities and calculating the invasion fitness from the game with the mutants.

\section{Connection with data}

Once we already have a selected model for a data-generating process, we may want to determine this model through the statistical procedure of parameter estimation. The input of the process is the data, measured from the physical systems of interest. The output of the process is a posterior probability function, the probability density of the estimated parameters under the model.\cite{Fearnhead2014InferenceApproximation, Komorowski2009BayesianApproximation}

The Bayes equation gives us the parameters' posterior distribution,
\begin{equation}
    P(H_k|D,I) = \frac{P(D|H_k,I)P(H_k|I)}{P(D|I)}.
\end{equation}

$D$ is a proposition asserting the data, coming from the data-generating process. The set of hypotheses $H_k$ will mean the following:

$H_k$ = \{The model $m_k$ with parameter values $\boldsymbol{\theta_i}$ is true\}.

And we write $H_k = M_k,\Theta_i$, with these new propositions meaning:

$M_k$ = \{The model $m_k$ is true\}.

$\Theta_i$ = \{The vector of parameters, $\boldsymbol{\theta}$, for the given model is between $\boldsymbol{\theta_i}$ and $\boldsymbol{\theta_i}+\boldsymbol{d\theta}$\}.

Being interested in parameter estimation, we work with a fixed model, so we may omit the proposition $M_k$ as always true for the model. We also omit the data probability, since it doesn't involve $\Theta_i$. Then, we work only with the estimation kernel, on the form
\begin{equation}
    P(\Theta_i|D,I) \propto P(D|\Theta_i,I)P(\Theta_i|I).
\end{equation}
There are two elements to deal with for the estimation, 1) the data likelihood given the parameter values, $P(D|\Theta_i,I)$, and 2) the parameter's prior information, $P(\Theta_i|I)$. Under absence of previous parameter information, the main function of the prior is to encode the parameters known properties, such as their domains. The likelihood is usually computed as a convolution between the model distribution and the error distribution (for a model with added errors).

Models of reaction networks have a straightforward statistical implementation, in theory. By somehow obtaining $\boldsymbol{n}$ (for a stochastic model) or $\boldsymbol{\eta}$ (for a deterministic approximated model), we can provide simple numerical solutions as a model for the data in order to feed a likelihood. The parameters we want to estimate in this case are the reaction rates and also unknown initial states or measurement errors.

The real problem is that we almost never can afford to solve the master equation for reaction networks, not even numerically. Out of the box, a numerical solution for parameter estimation would consist of simulating a large amount of samples for every relevant set of parameter values, which is a prohibitive computational effort. Then, the approximations to the master equation are, in principle, especially useful for this task. More standard stochastic approximations, such as the Langevin equation, unfortunately lead to hard obstacles in terms of matching the evolution of the model with the observed data, which demands complex fixes that are usually called observational bridge constructs. The SSE, on the other hand, can be readily used in the likelihood function, and this is a great feature of the expansion. The reason is that the SSE provides a full model for the evolution of the state distribution between observations. But even the estimation with LNA likelihood can be extremely demanding on computations.

Here, we will just explain and use the process of parameter estimation on reaction networks with the approximated model of the deterministic limit. Although it is a more simplistic approximation that loses all information on the structure of noise around the mean of the model (which is very useful for parameter estimation), we can draw on the benefits of it being easier to implement and also fairly fast. The deterministic model then shows the least the estimation procedure can do in this context.

Suppose we have a numerical solution for $\boldsymbol{\eta}$. Since the model is deterministic, we have the simple form of a likelihood where the model is its mean, and we can choose the most appropriate form of distribution for the observation noise. Since the LNA gives a structured Gaussian distribution of noise, dependent on $\boldsymbol{\eta}$, we will also choose a Gaussian with constant and diagonal covariance matrix, $\Sigma=diag(\sigma_i^2)$ as an approximation to the form of the likelihood; that can be seen as component observations with independent errors with constant variance. Then, the measurement model we will use is
\begin{equation}
    \boldsymbol{\eta} = \boldsymbol{x} + e,
\end{equation}
where $\boldsymbol{x}$ is the data and we are implicitly considering only the measured components. The error is then $e \sim \mathcal{N}(0,\Sigma)$. Setting the transition rates and other defining constants as the parameter vector $\boldsymbol{\theta}$, with dimension equal to the number of parameters to estimate, the likelihood from independent measurements becomes
\begin{equation}
    P(X|\Theta) = \prod_k\mathcal{N}(\boldsymbol{x_k}|\boldsymbol{\eta}_k(\boldsymbol{\theta},t_k),\Sigma),
\end{equation}
with $k=1,2...,K$ representing the data points, consisting of the pair $(t_k,\boldsymbol{x}_k)$. In order to initialize the model for $\boldsymbol{\eta}$, we choose the parameters $\boldsymbol{\eta}_0$ as initial states for a time right before the first measurement $t_0<t_1$, and also estimate the initial states. If we wanted to perform the estimation from the LNA model instead, the main difference would be to sum into $e$ the solution to the LNA.

In order to properly illustrate the parameter estimation process, we consider the Lotka-Volterra model:
\begin{gather}
    X_1 \xrightarrow{w} 2X_1 \nonumber\\
    X_1+X_2 \xrightarrow{\gamma} (1+\delta) X_2 \nonumber\\
    X_2 \xrightarrow{\mu} \emptyset.
\end{gather}

We generate data using the stochastic simulation algorithm on a medium-sized space of $\Omega=100$: from a sample trajectory, we extract $30$ measurements at random times for both preys and predators. For the parameter estimation, I used a Markov chain Monte Carlo algorithm with STAN.\cite{stanref} (Appendix 2)

We then use the deterministic dynamics as a statistical model of the data-generating process, thus the likelihood becomes a Gaussian having the model as the mean and the variance $\sigma^2$ as a proxy noise level also to be estimated. For simplicity, we use the generally well-suited exponential priors for scale parameters. The priors will reflect the order of magnitude of the parameters, with means equal to one of $0.1$, $1$, or $10$, depending on the parameter.

\begin{figure}[h]
    \centering
    \includegraphics[scale=1.55]{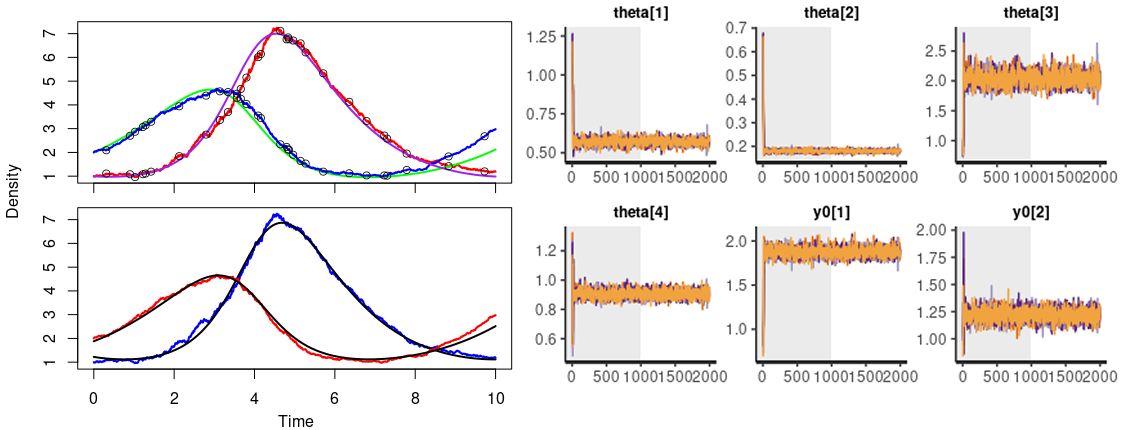}
    \caption{\textbf{Parameter estimation on the Lotka-Volterra model}. \textbf{Upper left:} Stochastic sample with 30 randomly extracted measurements for both preys and predators, compared with the deterministic model. \textbf{Lower left:} Same stochastic sample, compared with the deterministic curve generated with the mean estimated parameters. \textbf{Right:} Trace plots for posterior samples of the 4 network parameters and the initial state of the model. The gray region indicates the warm-up iterations.}
    
    \label{LVestimationFig}
\end{figure}

The statistical task is to estimate the parameters $\boldsymbol{\theta}=(w,\gamma,\delta,\mu, x_0, y_0, \sigma)$, where $(x_0,y_0)$ is the initial state of preys and predators used to initialize the statistical model, and $\sigma$ is the proxy standard deviation of the measurements. The initial state, for $t=0$, was chosen as $200$ preys and $100$ predators, meaning densities of $(x_0=2,y_0=1)$. Figure (\ref{LVestimationFig}) shows the sampled trajectory and measurements together with both the deterministic model and the mean estimated model, along with trace plots of the posterior. The table (\ref{tab:table1}) compares the estimated values with the real parameter values used for data generation.

The estimation process runs with 4 chains. For all chains, the trace plots indicate the expected behavior of the jumps, a "fuzzy caterpillar" shape, of well-mixed exploration of the posterior. Note that, with just the noisy extracted measurements and the statistical model, we are able to, in theory, estimate the particular place of the multi-dimensional parameter space in which the system operates.

\begin{center}
\begin{table}[]
\centering
\caption{\textbf{Estimated parameters.} The estimated posterior yields the mean values and standard deviations shown in the table, compared with the real parameter values. The standard deviation of the measurements $\sigma$ has no real value because it is a proxy of the actual noise levels coming from the stochastic process.}

\begin{tabular}{|c|c|c|}
\hline
\textbf{Parameter} & \textbf{Mean Estimation} & \textbf{Real Value}\\ \hline
$w$                  & $0.57 \pm 0.02$ & $0.55$ \\ \hline
$\gamma$                  & $0.18 \pm 0.01$ & $0.18$ \\ \hline
$\delta$                  & $2.04 \pm 0.11$ & $2.00$ \\ \hline
$\mu$                  & $0.91 \pm 0.03$ & $0.84$ \\ \hline
$x_0$                  & $1.88 \pm 0.05$ & $2.00$ \\ \hline
$y_0$                  & $1.22 \pm 0.05$ & $1.00$ \\ \hline
$\sigma$                  & $0.15 \pm 0.02$ & - \\ \hline

\end{tabular}

\label{tab:table1}
\end{table}
\end{center}

\section{Further improvements}

At the moment, beyond working on possible corrections and better explanations, I'm interested in three additions to the framework:

1) To write down a specific form of games I already worked out and implemented on a work that is currently under review, a model of mating dynamics and parental investment. The implementation involves considering games as a whole generation where individuals generate payoff instead of offspring, but they die, and the generation ends when there is no more active population. Then, using adaptive dynamics to calculate trajectories of trait evolution.

2) To devise a version of the framework with multiple systems of different sizes, connected with each other through the migration of individuals, as a way to model spatial structures. The instantaneous difference in densities makes it not trivial to model spatial structure just with the consideration of meta-populations (not sure if this worry is sound).

3) To think more about and test out the consideration of distribution of parameters affecting populations and reactions, in the way portrayed in Rose et al, 2021.\cite{Rose2021HeterogeneityModels} We define a continuous parameter $\epsilon$ representing the value of a variable which elements can possess (like a currency) according to a given distribution, then this variable influences the rate of some reactions according to $\propto \epsilon p(\epsilon)$, effectively distributing continuous reaction weights.

4) To use distributions of spatially-structured densities of each population to inform the interaction rates. Two individuals have a higher chance to interact (thus interact more) where their joint distribution of densities is higher.

5) To write down another generalization of transition rates to include spatial-less network models, making $\omega_r=\Omega$ so all individuals share the same space that is a unique system-size active cell for interactions, but then the transitions $W_r\propto w_{ij}$ are proportional to the edge weights of the network representing the strength of their interaction links.

\FloatBarrier
\newpage

\section{Appendix: Bayesian Derivation of a Master Equation for Markov Jump Processes}

This is a standard derivation of the master equation, but with fully Bayesian arguments. It is nice as a demonstration of Bayesian reasoning. Consider a system $\Gamma$, defined by the following assumptions:

1. $\Gamma$ exists in a discrete state space, with states that can be uniquely determined by a set of numbers, each describing a component of $\Gamma$ (usually translated to integer count numbers of each type of component). So, if $\Gamma$ is a system determined by two components, two species $N_1$ and $N_2$ with counts $n_1$ and $n_2$, then at a given time it's determined by the pair $(n_1,n_2)$ contained in the set of possible states. We denote the state of the system with the vector $\boldsymbol{n}$ with dimension equal to the number of system's components. In the example, $\boldsymbol{n}=(n_1,n_2)$.

2. $\Gamma$ evolves by changing states along a continuous passage of time. So, $\Gamma$ has a continuous set of time instants and is a "jump process".

3. $\Gamma$ obeys the Markovian property and we know the transition rates for the system. We'll rewrite them in terms of the transition probabilities. Also, the jumps to the many different states are independent events.

4. We can divide the time set into defined intervals $dt$ for which we can consider $\mathcal{O}(dt^2)/dt \to 0$ for $dt\to0$ and that we can assure transition rates to be approximately constant during $dt$.

So, if we know that $\Gamma$ is in a state $\boldsymbol{n_1}$ at a time $t_1$, it can jump to any other state $\boldsymbol{n_2}$ at a posterior time $t_2$ with a probability $P(\Gamma_{\boldsymbol{n_2},t_2}|\Gamma_{\boldsymbol{n_1},t_1},Z_{t_1,t_2})=Tr(n_1,t_1\to n_2,t_2)$, with $Z_{t_1,t_2}$ = \{There are no jumps during the interval $t_2-t_1$\} and $n_1 \neq n_2$. Since $\Gamma$ is Markovian, the transition probability does not depend on states before $t_1$. We define $\Gamma_{\boldsymbol{n},t}$ = \{$\Gamma$ is in state $\boldsymbol{n}$ at time $t$\}. In order to completely specify the system, we must connect the transition probabilities to the known transition rates. They are defined as follows:
\begin{equation}
    Tr(n_1,t_1\to n_2,t_2)=W_{\boldsymbol{n_1},t_1 \to \boldsymbol{n_2},t_2}dt,
\end{equation}
as long as $t_2-t_1=dt$. But we have problems. When the system jumps, this probability breaks; how many times can we expect it to jump during a time interval $dt$? Also, what is the probability of the system remaining in the same state after $dt$, the negation $\overline{\sum_{\boldsymbol{n_j}}Tr(n,t\to n_j,t+dt)}= 1-\sum_{\boldsymbol{n_j}}Tr(n,t\to n_j,t+dt)$? Can we know it?

\subsection*{Transitions}

We are interested in the limit $dt \to 0$, so we can solve our problems by proving the following statement: During a passage of time $dt$ starting at time $t$, the system can jump once, from state $\boldsymbol{n}$ to any different state $\boldsymbol{n_i}$ with probability $W_{\boldsymbol{n},t \to \boldsymbol{n_i},t+dt}dt$. Also, the system can jump more than once with probability $\mathcal{O}(dt^2)$ and remain in state $\boldsymbol{n}$ with probability $1-\sum_{n_i} W_{\boldsymbol{n},t \to \boldsymbol{n_2},t+dt}dt + \mathcal{O}(dt^2)$.

For that, consider the propositions, using the notation with implicit dependency of time $W_{n,n_j}=W_{\boldsymbol{n},t \to \boldsymbol{n_j},t+dt}$:

$K_k$ = \{With $\Gamma$ being in state $\boldsymbol{n}$ at time $t$, exactly $k>0$ transitions occur during the next interval $dt$, $k_j$ from $\boldsymbol{n}$ to $\boldsymbol{n_j} \neq \boldsymbol{n}$ with constant probability $W_{n,n_j}dt$ and the constraint $\sum_{n_j}k_j=k$.\}

With constant independent transitions, $P(K_k)$ follows a multinomial distribution with $k$ trials and a number of possible outcomes equal to the number of possible states. One of its possible outcomes never happens in any trial, representing the system jumping to nowhere in that trial, so:
\begin{equation}
    P(K_k) = \sum_{\sum k_j=k}\frac{k!}{\prod_j k_j!}\prod_{n_j \ne n} (W_{n,n_j}dt)^{k_j}(1-\sum_{n_j \ne n} W_{n,n_j}dt)^0 = (\sum_{\sum k_j=k}\frac{k!}{\prod_j k_j!}\prod_{n_j \ne n} W_{n,n_j}^{k_j})dt^k.
\end{equation}
We can see that this probability is proportional to $dt^k$, so we have $P(K_k) = \mathcal{O}(dt^{k})$. In particular, 
\begin{equation}
    P(K_1) = \sum_{n_j \ne n} W_{n,n_j}dt.
\end{equation}
For no transitions, we have the proposition $K_0$, written as a negation $K_0 = \overline{\sum_k K_k}=\overline{\sum_{\boldsymbol{n_j}}Tr(n,t\to n_j,t+dt)}$. Noting that the $K_k$s are mutually exclusive, using the sum rule, we have
\begin{equation*}
    P(K_0) = P(\overline{\sum_k K_k}) = 1- P(\sum_k K_k) = 1 - \sum_kP(K_k) =
\end{equation*}  
\begin{equation}
    1-P(K_1) - \sum_{k>1}P(K_k) = 1-\sum_{n_j \ne n} W_{n,n_j}dt + \mathcal{O}(dt).
\end{equation}
This ends our justification and solves our problems. We can now make sure that at most one transition occurs during $dt$ in the limit.

\subsection*{Master Equation}

Finally, we turn to the task of building the master equation. Let's give some easier names to our relevant propositions:

$X_0$ = $\Gamma_{\boldsymbol{n_0},t_0}$ = \{$\Gamma$ starts in an initial state $\boldsymbol{n_0}$ at time $t_0$\}.

$X$ = $\Gamma_{\boldsymbol{n},t}$ = \{$\Gamma$ is in state $\boldsymbol{n}$ at time $t>t_0$\}.

And for each possible state $\boldsymbol{n}_i$:

$Y_i$ = $\Gamma_{\boldsymbol{n_i},t}$ = \{$\Gamma$ is in state $\boldsymbol{n_i}$ at time $t'<t$ with $t'>t_0$\}.

The goal now is to assign a probability to proposition $X$ using the $Y_i$s. Let's look at the proposition $\sum_i Y_i$; it means $Y_1$, or $Y_2$, or $Y_3$, etc. It essentially means that $\Gamma$ is in any possible state at time $t'$, and this is always true, the set $\{Y_i\}$ is exhaustive. Also note that $Y_i$s are mutually exclusive, because at the same time $\Gamma$ can only be in one state. So the set $\{Y_i\}$ is a partition of the event space at time $t'$, a set of mutually exclusive events covering the whole space. Since the sum of $Y_i$s is always true, and using product properties, we can write
\begin{equation}
    X = (\sum_i Y_i),X = \sum_i X,Y_i.
\end{equation}
Let's begin assigning probabilities to porpositions. Note that the products $X,Y_i$ are also mutually exclusive, so we have, using the sum rule
\begin{equation}
    P(X|X_0) = P(\sum_i X,Y_i|X_0) = \sum_i P(X,Y_i|X_0).
\end{equation}
Now we use the product rule
\begin{equation}
    P(X|X_0) = \sum_i P(X|Y_i,X_0)P(Y_i|X_0)
\end{equation}
and then the Markovian property, that says $P(X|Y_i,X_0)=P(X|Y_i)$,
\begin{equation}
    P(X|X_0) = \sum_i P(X|Y_i)P(Y_i|X_0).
\end{equation}
See that all this is just the law of total probability applied to $X$ with the partition $\{Y_i\}$. Now, why is it relevant to rewrite $P(X|X_0)$ in terms of the $Y_i$s? It is because, with our specification of $\Gamma$, we have knowledge about local transition probabilities, but the known initial state $X_0$ may be as far as we wish from the arbitrary state $X$ we want to describe. Using the $Y_i$s as bridges, we can make $t'$ "adjacent" to $t$ and smuggle the known transition probabilities into our derivation. With adjacent meaning distant by an interval $dt$.

We need to specify a $t'$ of $Y_i$ that is adjacent to the $t$ of $X$: $t'= t-dt$. If this is true, we have the probabilities $P(X|Y_i)$ in terms of the transition rates. There are two cases; 1) $\boldsymbol{n_i}=\boldsymbol{n}$ and it means that no transitions occur, and 2) $\boldsymbol{n_i} \ne \boldsymbol{n}$ and it means that some transition with rate $W_{\boldsymbol{n_i},t-dt \to \boldsymbol{n},t}$ occurs. So we separate the sum into these two possibilities
\begin{equation}
    P(X|X_0) = \sum_{\boldsymbol{n_i} \ne \boldsymbol{n}} P(X|Y_i)P(Y_i|X_0)+P(X|Y_n)P(Y_n|X_0),
\end{equation}
with $Y_n$ defined as $Y_i$ for the case of $\boldsymbol{n_i}=\boldsymbol{n}$. The transition probabilities are, using the same implicit time-dependency notation as above,
\begin{equation}
    P(X|Y_i) = W_{\boldsymbol{n_i},\boldsymbol{n}}dt + \mathcal{O}(dt^2)
\end{equation}
because we are going from $\boldsymbol{n_i}$ to $\boldsymbol{n}$. The probability of no transition is
\begin{equation}
    P(X|Y_n) = 1-\sum_{\boldsymbol{n_i} \ne \boldsymbol{n}} W_{\boldsymbol{n},\boldsymbol{n_i}}dt + \mathcal{O}(dt^2)
\end{equation}
because we are going from $\boldsymbol{n}$ to all other $\boldsymbol{n_i}$s. Note the exchange in the indexes of $W$. Putting more clearly, in case 1 the system is jumping from $\boldsymbol{n_i}$ to $\boldsymbol{n}$, and in case 2 the system already is in $\boldsymbol{n}$ and we consider the negation of it going to any other possible $\boldsymbol{n_i}$.

Inserting in the equation for $P(X|X_0)$, we have
\begin{equation*}
    P(X|X_0) = \sum_{\boldsymbol{n_i} \ne \boldsymbol{n}} \left(W_{\boldsymbol{n_i},\boldsymbol{n}}dt + \mathcal{O}(dt^2)\right)P(Y_i|X_0)+
\end{equation*}
\begin{equation}
    \left(1-\sum_{\boldsymbol{n_i} \ne \boldsymbol{n}} W_{\boldsymbol{n},\boldsymbol{n_i}}dt + \mathcal{O}(dt^2)\right)P(Y_n|X_0).
\end{equation}
Just reorganizing the equation, we arrive at
\begin{equation}
    \frac{P(X|X_0)-P(Y_n|X_0)}{dt} = \sum_{\boldsymbol{n_i} \ne \boldsymbol{n}}\left(W_{\mathbf{n_i},\boldsymbol{n}}P(Y_i|X_0)-W_{\boldsymbol{n},\boldsymbol{n_i}}P(Y_n|X_0)\right)+\frac{\mathcal{O}(dt^2)}{dt}.
\end{equation}
Finally, we perform the limit $dt \to 0$. With this, the left side of the equation becomes the derivative of $P(X|X_0)$ in relation to time and $\frac{\mathcal{O}(dt^2)}{dt} \to 0$. $P(Y_n|X_0)$ on the right side becomes $P(X|X_0)$ as $t' \to t$ (note that the $Y_i$s now represent $\Gamma$ in time $t$ with the limit imposing $t' \to t$). We have the Master Equation:
\begin{equation}
    \frac{d P(X|X_0)}{d t} = \sum_{\boldsymbol{n_i} \ne \mathbf{n}}\left(W_{\boldsymbol{n_i},\boldsymbol{n}}P(Y_i|\{t'=t\},X_0)-W_{\boldsymbol{n},\boldsymbol{n_i}}P(X|X_0)\right).
\end{equation}
We can now change the probabilities to the more explicit distributions notation. The distribution that $P(X|X_0)$ follows has as variables the state vector $\boldsymbol{n}$ and the time $t$. If we define the probability of no transitions occurring as $W_{\boldsymbol{n},\boldsymbol{n}}$, we can sum over all states of $\Gamma$ without altering the equation (note that the additional term $n_i=n$ ends up being zero). Calling the distribution $P(X|X_0)=\Pi(\boldsymbol{n},t)$, we have
\begin{equation*}
    \frac{d \Pi(\boldsymbol{n},t)}{d t} = \sum_{\boldsymbol{n_i}}\left(W_{\boldsymbol{n_i},\boldsymbol{n}}\Pi(\boldsymbol{n_i},t)-W_{\boldsymbol{n},\boldsymbol{n_i}}\Pi(\boldsymbol{n},t)\right),
\end{equation*}
\begin{equation}
    \Pi(\boldsymbol{n},t_0) = \delta (\boldsymbol{n},\boldsymbol{n_0}).
    \label{masterequation1}
\end{equation}
Note that we can generalize the proposition $X_0$ into a set of propositions to mean that the state of $\Gamma$ in $t_0$ is uncertain, with different probabilities of being in different states. We don't need to know the exact initial state for the equation to be valid. For systems with a finite number of states, we can even know nothing about the initial state, assigning to the set of $X_0$ a uniform probability distribution over the sates.

The solution of this equation gives the probability of proposition $X$ happening once that $X_0$ happened, that means $\Gamma$ has transitioned to state $\boldsymbol{n}$ after an arbitrary number of jumps during an arbitrary time interval $t-t_0$.

We can interpret the Master Equation in terms of gains and losses in probability; it means that the right side is viewed as a net gain in probability at time $t$, the first term being the gain from transitions into $\boldsymbol{n}$ and the second term being the loss from transitions away from $\boldsymbol{n}$.

The Master Equation is the differential form of the Chapman-Kolmogorov equation.

\newpage
\section{Appendix: Introduction to MCMC}

In the parameter estimation process, once the model is ready, we are in theory expected to integrate the kernel of the posterior for every set of values in the multidimensional parameter space; then, in order to extract information from the posterior we have to marginalize and calculate expectations through more integration on the posterior. Our task of estimating parameters transforms into a computational burden of integrating functions on a high dimensional space and which normally feature a slim geometry of probability mass, making integration especially painful. For this reason, direct integration is virtually never a viable option in the Bayesian analysis. One of the main methods to determine probability distributions and widely used in Bayesian parameter inference is the Markov Chain Monte Carlo (MCMC). Here, I present an introduction to MCMC in my own words.

\subsection*{Monte Carlo}

A Monte Carlo method is one that in general transforms samples into integrals. This is built upon the law of large numbers, that basically shows us how to view uncertain events as certain events plus an approximation error.

We'll work out the intuitions through one dimensional continuous objects, but they can readily be generalized to more dimensions and discrete spaces. Suppose a data generating process $\{X_i^p\}$ = \{The variable $x$ modeled by the probability density $p(x)$ is in [$x_i,x_i+dx$]\}, with probability $P(X_i^p)=p(x_i)dx$. According to the law of large numbers, we can calculate the mean of any function $f(x)$ over a density $p(x)$ by using a set of $N$ samples as the approximation
\begin{equation}
    \langle f \rangle = \int f(x) p(x) dx \approx \frac{1}{N}\sum_{i=1}^N f(x_i),
    \label{montecarlo}
\end{equation}
an unbiased estimation with error $\mathcal{O}(N^{-1/2})$. As a particular well-known case, we have $\langle x \rangle \approx \sum_i x_i/N$, called the sample mean. But then, if we view $f(x)p(x)$ as a simple function of a real variable $x$, this is actually a method for calculating the definite integral of $f(x)p(x)$ over a support set through the sample mean. So the law of large numbers can act as a connection between samples of distributions and deterministic integrals. In particular, for a uniform density over an interval of length $L$, we have $p(x)=L^{-1}$, and
\begin{equation}
    \int_L f(x) dx \approx \frac{L}{N}\sum_{i=1}^Nf(x_i).
    \label{montecarlo2}
\end{equation}
In this case, we use the uniform samples as a sort of "mining" of function values that in the limit will equally distribute themselves around the function mean. And if we map the area under the curve of $f(x)$ into a rectangle by an area-preserving transformation, that rectangle would have a length of $L$ and a height of $\langle f(x) \rangle$.

By using monte carlo integration, we can focus on just sampling the posterior. It is a much easier task than determining a posterior, marginalizing it, and calculating expectations.

\subsection*{Importance Sampling}

The method in Eq. (\ref{montecarlo2}) presupposes that we draw samples from the distribution $p(x)$, but we may need or want to draw samples from another distribution $q(x)$, for example the standard case of drawing from uniform distributions in algorithms. Then, it would be useful if we could input the sampling from a different distribution $q(x)$ into calculations for $p(x)$. This can be done as the trick
\begin{equation}
     \langle f \rangle = \int f(x) \frac{p(x)}{q(x)} q(x) dx \approx \frac{1}{N}\sum_{i=1}^N f(x_i^q) \frac{p(x_i^q)}{q(x_i^q)}.
\end{equation}
So it is the same as sampling the function $f(x)p(x)/q(x)$ from the $q(x)$ distribution. In this context, we can say that we are giving to each $x_i^q$ an importance, or weight, of $p(x)/q(x)$ in order to calculate the mean of $f(x)$ under $p(x)$.

All this is, in principle, of great value for the parameter estimation process through the posterior distribution. With it, we may sample parameters from the posterior in order to calculate estimators for them, such as the mean, even if we have to sample primarily from another distribution. But then we run into a problem: we can have at most the kernel of the posterior, not the entire density. So, we have the posterior represented by the density $p^*(x)=k(x)/Z$, where $k(x)$ is the kernel and $Z=\int_L k(x)dx$ is the unknown normalization factor of the posterior. But then, since $Z$ is actually an integral, there is now a straightforward way to calculate it:
\begin{equation}
    Z = \int_L k(x)dx \approx \frac{L}{N}\sum_i \frac{k(x_i^q)}{q(x_i^q)}.
\end{equation}
Thus, by estimating $Z$ itself, we can distribute importance (weight) to values of $x$ in the interval according to an estimated density from the known kernel.

\subsection*{Rejection Sampling}

But then, we notice that calculating from narrow distributions by sampling other densities like that may be an inefficient process. If $p(x)$ and $q(x)$ don't match, many samples $x_i$ can have a negligible importance in relation to contributing to the probability mass, especially in high dimensional spaces. That's because the probability mass of kernels is usually concentrated in a narrow subset of the parameter space (called typical set), and it gets more concentrated for higher dimensions. This mismatch is the price we pay in order to sample from a distribution using another distribution. In an estimation task, if we could sample the $x_i$ from the posterior itself, it would be a much more efficient sampling process, optimally efficient in this sense. A way to do this is to reject some $x_i^q$ on the basis of their importance under the kernel. It makes the sample generation less computationally efficient to assure efficiency in the convergence of the integration. This transforms the sampling under $q(x)$ into a proposal of sampling, and a candidate sample is filtered under $p(x)$ (or the kernel). For us, a major advantage of this method is that we don't need to estimate $Z$, which is a much more inefficient process. We'll see that the acceptance-rejection of $x_i^q$ can be defined with only the kernel.

For each sampled $x_i^q$, we draw a uniform $u$ in the interval $[0,1]$, and we accept $x_i^q$ if
\begin{equation}
    u < \frac{k(x_i^q)/q(x_i^q)}{max[k(x)/q(x)]},
\end{equation}
intuitively meaning that, in order to be accepted, $x_i^q$ must fall under the curve of $k/q$. Thus, we reject the sample if it falls off the curve of the kernel, in a region defined by the constant boundary $max[k(x)/q(x)]$ that ensures to encapsulate the whole curve of $k/q$. This boundary (and also $q$) is of course considered only from values inside the support of the kernel. Note that, by using a ratio as the filtering criterion, we don't need information of $Z$ (it is only a scale on the kernel). In order to justify this, consider the propositions:

$A$ = \{A value was accepted\},

$X$ = \{The sampled value is $x$\}.

Then, $P(X|A)=P(A|X)P(X)/P(A)$  has a density
\begin{equation}
    p(x|A)=\frac{(k/(mq))q}{P(A)}=\frac{k(x)}{mP(A)},
\end{equation}
where we defined $m=max[k(x)/q(x)]$. $P(A)$, the probability of a proposal being accepted, irrespective of its value, can be calculated by marginalizing $P(X,A)$ over $x$:
\begin{equation}
P(A) = \int P(X,A)dx = \int P(X|A)P(X)dx = \int \frac{k(x)}{mq(x)}q(x)dx = \frac{Z}{m}.
\end{equation}
Then, $p(x|A)= k(x)/Z$, and the accepted samples are distributed according to the desired density, in our case the posterior $p^*(x)$.

The most widespread picture of a monte carlo integration is done with rejection sampling. Instead of directly calculating the integral Eq. (\ref{montecarlo2}) from a uniform sampling, the uniform sampling is used as a proposal. Then, the function $f(x)$ itself is used as a kernel for the rejection-acceptance step. The simple integral then equates with the monte carlo estimation of $Z$. The visualization of this process is one of dots accumulating both inside and outside the curve of $f(x)$; the dots falling inside the function are the accepted ones, and those falling outside are rejected.

\subsection*{Markov Chain}

The task of determining a posterior distribution is one of finding its probability mass in the parameter space. We saw that a rejection sampling technique can assure that sampling will efficiently represent the posterior probability mass. But we just shifted the problem to the burden of proposing sample candidates. In high dimension parameter spaces, the probability mass will represent just a slim proportion of the space's volume. This means that a lot of proposals will get rejected if our choice of $Q(X)$ isn't already aligned with the kernel. Thus, we are still left with the pressing goal of electing an efficient proposal distribution, one that listens to the location of the posterior's probability mass.

The idea is to use the posterior's geometry in order to devise a criterion. In general, the probability mass is not scattered over the parameter space, but packed inside a specific typical set. We may guess that the typical set is concentrated around the mode, as is the case of the geometry of a one dimensional Gaussian distribution. But at higher dimensions, it non-trivially spreads away from the mode; because despite the importance of points is decreased, the volume of the typical set increases in regions away from the mode. Thus, the geometry of the high dimension posterior in general resembles a narrow band around the region of large importance. We must devise a sampling method that probes the parameter space for this set and then wanders over it with good mixing.

This suggests that we correlate the sampling process, in an attempt to encode the goal of "getting closer" to the typical set once a sample falls far away from it. More formally, we want, given a sample, to distribute the next sample in a way that actively searches for probability mass. With that, $q(x)$ will shape itself according to the geometry of the posterior, granting a sufficiently high acceptance of proposals. For example, simply proposing samples that are nearby an accepted sample already does wonders in increasing the chances of acceptance, because we can expect that accepted samples are more probably located in good neighborhoods (the posterior mass is not scattered over the parameter space, but concentrated). In other words, if a sample is accepted, there is a higher chance that it is closer to the typical set than rejected samples, because the importance for acceptance is based on the kernel itself.

But if we want to correlate a sample with the previous sample, we want to make the sampling process into a Markovian chain. And since we want to lock it as being distributed as the posterior, it must be in equilibrium. Then the problem is reduced to the coordination of a proposal and an acceptance that result in both the equilibrium state of a Markov chain and the posterior distribution. In theory, no matter where the sampling process starts, it can converge to an equilibrium that mimics the sampling of the posterior. Since we now incur in the drawback of having correlated samples, we must ensure a good sampling mixture in order to use the process for estimations (ensure that the process is really able to capture the whole target distribution, and does not "get stuck" in certain regions).

Another problem to consider is that the acceptance process is not that well defined yet, because the determination of a quantity like $m=max(k/q)$ already is an optimization problem. The idea of probing for the typical set from a current sample can also be used to address this and devise a local acceptance criterion. 

This process of sampling from a Markov chain in order to calculate expectations from a desired target distribution is what is generally called a Markov chain Monte Carlo sampler (MCMC).

\subsection*{Metropolis-Hastings}

The algorithm of Metropolis-Hastings is a MCMC sampler built on a property of reversible Markov chains, an equilibrium constraint called detailed balance. Consider the set of statements about a chain at equilibrium \{$X_i^{(t)}$\} = \{The state of the chain $x$ is in [$x_i,x_i+dx$] at time $t$\}. Then, in detailed balance,
\begin{equation}
    P(X_i^{(t-1)})P(X_j^{(t)}|X_i^{(t-1)}) = P(X_j^{(t-1)})P(X_i^{(t)}|X_j^{(t-1)}),
\end{equation}
noting that $P(X_i^{(t-1)})=P(X_i)$, because it is at the equilibrium. This is the same as saying that $P(X_i^{(t-1)},X_j^{(t)})=P(X_j^{(t-1)},X_i^{(t)})$. Under detailed balance, the probability flux of the jump from $i$ to $j$ is the same as for the jump from $j$ to $i$, so there is no net flux in the chain; the transitions are pairwise in equilibrium. When we define a particular chain through its transition probabilities, if we make sure that the chain satisfies detailed balance with the posterior, then if it is a proper posterior, that is the unique equilibrium distribution of the chain. Thus, the requirement is to choose transitions satisfying
\begin{equation}
    \frac{P(X_j^{(t)}|X_i^{(t-1)})}{P(X_i^{(t)}|X_j^{(t-1)})} = \frac{k(x_i)}{k(x_j)},
\end{equation}
where $k(x)$ is the kernel of the posterior. The transition is the product of a proposal and an acceptance given proposal steps, so we must have
\begin{equation}
    P(X_j^{(t)}|X_i^{(t-1)}) = q(x_i,x_j)P(A_{ij}),
\end{equation}
where $q(x_i,x_j)$ is the sampling distribution, now dependent on both $x_i$ and $x_j$, and $A_{ij}$ = \{Given a proposal from $x_i$ to $x_j$, the jump to $x_j$ is accepted\}. This results in
\begin{equation}
    \frac{P(A_{ij})}{P(A_{ji})} = \frac{k(x_i)q(x_j,x_i)}{k(x_j)q(x_i,x_j)}.
\end{equation}
If we chose
\begin{equation}
    P(A_{ij}) = min \Bigg(1, \frac{k(x_i)q(x_j,x_i)}{k(x_j)q(x_i,x_j)}\Bigg),
\end{equation}
then it is a valid distribution for which the condition is always satisfied. 

The choice of a sampling proposal distribution $q(x_i,x_j)$ influences the speed of convergence of the chain. The particular Metropolis algorithm chooses it to be symmetrical (and making the acceptance independent of $q$), $q(x_i,x_j)=q(x_j,x_i)$, often a Gaussian $q(x_i,x_j) = \mathcal{N}(x_j|x_i,\sigma^2)$. In this case, the deviation $\sigma$ regulates a step-size for proposals, that can't be too large so as to miss the regions of interest and cause a large rejection rate or too small so as to be slow on convergence and mixing.

For a multi-dimensional parameter space, there is also a choice involved in the jumps being sequential on each dimension or in form of a batch update (updating all dimensions at once is more efficient). The samples taken before convergence are discarded in the estimation process (the initial samples are called warm up), and parallel exploration with multiple chains is advisable. There are actually many details to address in the practical use of MCMC to carry out the estimation process and also the diagnosis and analysis processes following it. We then want to rely on a good software that can take care of much of the engineering bits.

\printbibliography

\end{document}